\newcounter{myctr}

\documentclass{ws-acs}
\usepackage{url}
\usepackage{color}
\usepackage{ulem}

\begin{document}

\makeatletter
\def\@biblabel#1{[#1]}
\makeatother

\markboth{Alina S\^irbu, Vittorio Loreto, Vito DP Servedio, Francesca Tria}{Cohesion, consensus and extreme information in opinion dynamics}

%
\catchline{}{}{}{}{}
%

\title{Cohesion, consensus and extreme information in opinion dynamics }

\author{\footnotesize Alina S\^irbu}

\address{ Complex Networks and Systems Lagrange Laboratory, Institute for Scientific Interchange Foundation\\
Via Alassio 11/c, Turin, 10126,
Italy\\
alina.sirbu@isi.it}

\author{\footnotesize Vittorio Loreto}

\address{Sapienza University of Rome, Physics Dept.\\ P.le~A.~Moro 2,
        00185 Rome, Italy\\
 Complex Networks and Systems Lagrange Laboratory, Institute for Scientific Interchange Foundation\\
Via Alassio 11/c, Turin, 10126,
Italy\\
vittorio.loreto@roma1.infn.it}

\author{Vito D.~P.~Servedio}

\address{Sapienza University of Rome, Physics Dept.\\ P.le~A.~Moro 2,
        00185 Rome, Italy\\
vito.servedio@roma1.infn.it}

\author{\footnotesize Francesca Tria}

\address{ Complex Networks and Systems Lagrange Laboratory, Institute for Scientific Interchange Foundation\\
Via Alassio 11/c, Turin, 10126,
Italy\\
tria@isi.it}

\maketitle

\begin{history}
\received{(received date)}
\revised{(revised date)}
\end{history}

\begin{abstract}
Opinion formation is an important element of social dynamics. It has been widely studied in the last years with tools from physics, mathematics and computer science. Here, a \emph{continuous} model of opinion dynamics for \emph{multiple possible choices} is analysed. Its main features are the inclusion of \emph{disagreement} and possibility of \emph{modulating information}, both from one and \emph{multiple sources}. The interest is in identifying the effect of the initial cohesion of the population, the interplay between cohesion and information extremism, and the effect of using multiple sources of information that can influence the system. Final consensus, especially with external information, depends highly on these factors, as numerical simulations show. When no information is present, consensus or segregation is determined by the initial cohesion of the population. Interestingly, when only one source of information is present, consensus can be obtained, in general, only when this is extremely mild, i.e.\ there is not a single opinion strongly promoted, or in the special case of a large initial cohesion and low information exposure. On the contrary, when multiple information sources are allowed,  consensus can emerge with an information source even when this is not extremely mild, i.e.\ it carries a strong message, for a large range of initial conditions.

\end{abstract}


\section{Introduction}
Choices made during everyday life are the outcome of holding specific opinions on various subjects. These opinions drive human behaviour, and are at the basis of the existence of cultural groups when opinions are similar between individuals, or can also result in conflict when opinions clash on some topics. Subjects on which an opinion can be held, range from simple day to day choices, such as what to buy in the supermarket, or what phone company to choose, to more important questions such as religious and political views. Opinion formation and evolution is a very complex process influenced by several factors. The culture and personal predisposition are very important, together with human interaction, which makes opinions spread and form clusters. Additionally, external sources of information are introduced in society continuously, through mass-media or advertising campaigns.

Given its importance in society, the process of opinion formation has been widely studied, starting with social sciences and moving to physics and computational methods, which have proven to be very useful for this analysis. The emergence of new technologies and online social networks has changed both the manner of human interaction, but also the means to analyse it, by making available data describing its different aspects. Several modelling approaches for opinion formation based on interaction between individuals have been introduced. Very simple models are discrete spin-like models, where two opinion choices are represented as up or down spins.  Social interaction can be implemented by means of different types of rules: pairwise interaction (voter model, e.g. \cite{Zschaler2011}), local majority \cite{Galam2008a}, social impact \cite{Sznajd-Weron2001,Lewenstein1992,Nowak1996,vallacher2008}. This body of studies has enabled important observations on the mechanisms that drive opinion formation, and has been applied to describe elections, strikes, dynamics of mobile markets, changes in the number of privately owned companies, financial crises and culture formation \cite{Galam2010,Galam2011,Lima2012,nowak2000,Axelrod1997}.

Although important insights have been obtained from discrete models, sometimes opinions can be better represented by continuous variables. These would represent not only the choice made by individuals, but also the trust in that choice. Also, these could be used to model resource allocation decisions. Examples of such approaches are the Deffuant-Weisbuch \cite{Deffuant2000,Weisbuch2002} and Hegselmann-Krause \cite{Hegselmann2002} models, studying continuous opinions for two choices, while multiple dimensions are analysed in \cite{Lorenz2007,Fortunato2005,Deffuant2012,Lorenz2008}. The Continuous Opinions and Discrete Actions approach (CODA) \cite{Martins2008a} analyses internal probabilities for two or three discrete choices. 

Attractive dynamics have been considered by most previous models as the main factor driving opinion formation. In reality, however, individuals do not always agree and change their opinions to resemble those of their neighbours, but disagreement is very important in society \cite{Huckfeldt2004}. Some of the models above have been extended to include disagreement \cite{Radillo-Diaz2009,VazMartins2010,Kondrat2010,Sznajd-Weron2011,Nyczka2012,Kurmyshev2011,Hong2011,Acemoglu2010}. Furthermore, peer interaction is not the only feature driving opinion formation, since society is subject to external effects coming for example from mass-media. Some modelling approaches do consider this effect \cite{Carletti2006,Gargiulo2008a,Peres2011,VazMartins2010,Crokidakis2011a,Hegselmann2006}. However, for continuous multidimensional models these two elements (disagreement and external information) have not been studied to date.

Here, we discuss a novel model of opinion formation, which applies to situations when there are multiple opinion choices for one subject. The model uses continuous variables to represent the internal probability of an individual to select one of the possible choices. It is important to note that this is not an extension of the Deffuant model for vectorial opinions, where each continuous variable represented the opinion on a different topic \cite{Fortunato2005,Deffuant2012,Lorenz2008}. The approach presented here includes disagreement dynamics based on similarity between individuals and allows for the existence of external information, which can be modulated to account for mild or extreme messages. Importantly, one or more sources of information are introduced. An initial analysis of this modelling approach has been presented in \cite{sirbu2012}, for the model with one information source. Here we analyse the model further, to assess in more detail the interplay between the cohesion of the initial population and the number of opinion choices and information extremism. Additionally, an extension to multiple information sources is presented, which enriches the dynamics making possible the existence of population states unreachable by one information source only.

\section{Methods}
This paper analyses a model of opinion formation with disagreement and modulated information, introduced first in \cite{sirbu2012}. This is extended to allow for multiple sources of information, a critical feature when simulating some real situations. In the following, we briefly describe the original model, with one source of information, and we provide details on the extended version.
 
\subsection{Model with disagreement}
The model considers the situation when a choice has to be made between several ($K$) discrete options, such as choosing a telephone company, or voting for a single party at political elections. Each individual in a population of $N$ is described by an array of probabilities to choose a specific option (modelling the internal decision process): $\vec x=[p_1,p_2,\ldots,p_K]$. Since these are probabilities, each individual is represented by an element in the $K-1$ simplex $\sum_{k=1}^{K}{p_k}=1$. Although the model uses continuous opinions, it is important to note the difference from other vectorial continuous models for opinion dynamics (such as \cite{Fortunato2005,Deffuant2012}). In their case, each position in the vector is independent from the others, and represents the opinion of an agent on different subjects (for instance, one position could be the opinion on telephone companies, another on choice of supermarkets). In our case, however, all positions in the vector refer to the same subject, and give a weight to each of the possible opinion choices. 

\begin{figure}[th]
\centerline{\psfig{file=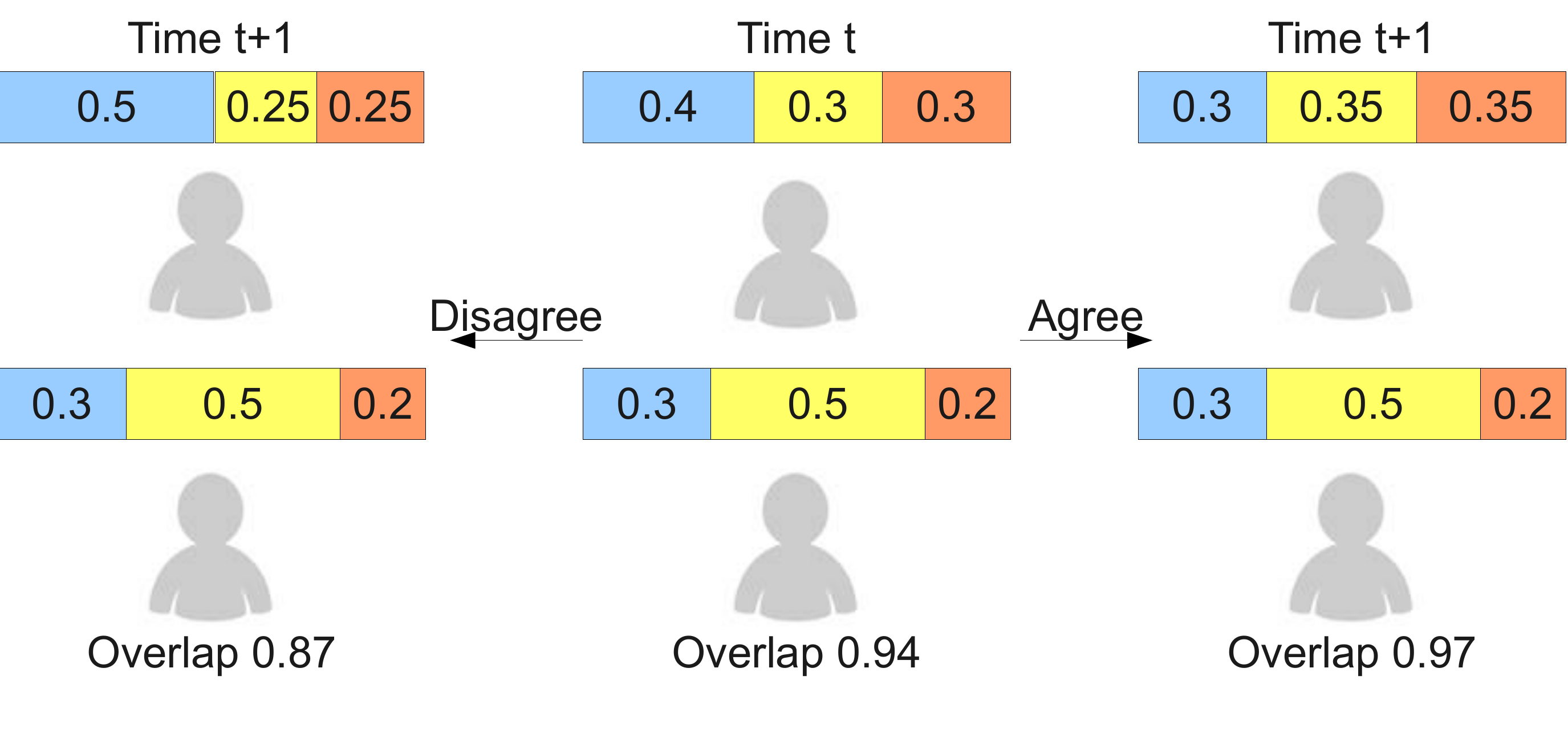,width=12cm}}
\caption{Interaction example with $K=3$, $\alpha=0.1$, $l=1$ (refer to the text for the parameters definition). Two individuals interact, holding opinions $[0.4,0.3,0.3]$ and $[0.3,0.5,0.2]$ and having at time $t$ overlap $0.94$ (middle column). The first one plays the role of the listener, while the other one is the speaker. In the case they agree (right column), the listener decreases the first opinion by $\alpha$, getting closer to the speaker opinion, and the other opinions are increased by $\frac{\alpha}{K-1}$. This leads to an increase in the overlap to $0.97$. Similarly, disagreement (left column) leads to an increase in the first position of the opinion vector of the listener, and a decrease in the other positions in order to maintain the normalization, resulting in a lower overlap at time $t+1$ ($0.87$). In the case where the uniform adjusting is not possible (opinions cannot have negative values or values greater than one), the difference is uniformly redistributed to non-negative positions until all the difference is accounted for.}\label{interaction}
\end{figure}

The agents in the population are connected by a social network, which we consider a complete graph (so to model a mean-field situation), and they influence each other either by agreeing or disagreeing with neighbours. A similarity measure between neighbours $i$ and $j$ is defined as the cosine overlap:
\begin{equation}
  o^{ij}=
    \frac{\vec{x}^{\,i}\cdot\vec{x}^{\,j}}{|\vec{x}^{\,i}| |\vec{x}^{\,j}|} =
    \frac{\sum_{k=1}^{K}{p_k^i p_k^j}}{\sqrt{\sum_{k=1}^{K}{(p_k^i)^2} \sum_{k=1}^{K}{(p_k^j)^2}}}.
\end{equation}
This takes values between 0 and 1, with 1 indicating complete agreement between agents while 0 indicates very different opinions. The overlap value is used to decide the type of interaction happening between  
$i$ and $j$, by computing a probability to agree or disagree:
\begin{equation}
  p_\mathrm{agree}^{ij}=\min(1,\max(0,o^{ij}\pm\epsilon)) \label{eq:agree},
  \end{equation}
  \begin{equation}
    p_\mathrm{disagree}^{ij}=1-p_\mathrm{agree}^{ij} \label{eq:disagree}.
  \end{equation}
  Here, $\epsilon$ is a noise term that allows for a small probability to agree 
when $o^{ij}=0$ and to disagree when $o^{ij}=1$.
  Agreement results in one of the individuals, the listener, changing a random position ($l$) in the opinion vector in the direction of the other, while through disagreement the change leads to a larger difference between agents:
   \begin{equation}
	p^i_l(t+1)=
	\begin{cases}
		p_l^i(t) \pm \alpha\, \mathrm{sign}(p_l^j-p_l^i) & \text{if } |p_l^j-p_l^i|>\alpha
		\\ 
		p_l^i(t) \pm \frac{1}{2}(p_l^j-p_l^j) & \text{otherwise.}
	\end{cases}
\label{interEq}
\end{equation}
with plus occurring for agreement and minus for disagreement. Hence, the listener changes position $l$ by a fixed step $\alpha$, as long as the difference between the two opinions is not smaller than $\alpha$, when half the difference is used for the change. The other opinions are adjusted uniformly to maintain the unit sum. Figure \ref{interaction} exemplifies the interaction rule.

\subsection{One external information source}
The system described above can be influenced by mass-media or other external sources, which are introduced in the model as a static agent $\vec I=[I_1,I_2,\ldots,I_K]$ with $\sum_{k=1}^{K}{I_k}=1$. At each time step each individual interacts with a randomly chosen neighbour and then, with probability $p_I$, with the external information. Hence interaction with the information does not imply less peer communication. Interaction with $\vec{I}$ follows the same rules as that with a peer (the individual being of course the listener).

Previous models of opinion dynamics introduced an external information source in a similar manner, through a static external agent (e.g.\  Axelrod, Sznajd). In these cases, due to the discrete opinion values, the external message promoted one option only, out of all possibilities. In our model, this would translate into using $\vec{I}=[0,\ldots,1,\ldots,0]$. However, here, multiple options can be also promoted by the external information, by using non-null values for more elements of $\vec{I}$. In this way, information can be modulated to account for the real situation, making possible extreme (when one of the elements of $\vec{I}$ is very close to 1 and the rest almost null) and mild (when more elements of $\vec{I}$ are non null) information. A similar possibility was given for models like Deffuant, for two choices only: considering that the opinion value is the strength of a choice over the other, external information far from the boundaries of the opinion interval can be considered mild.

\subsection{Multiple external information sources}\label{sec:MI}
Using one external source of information means aggregating all mass-media effects into one value. The possibility of modulating $\vec{I}$ is very important in this case, since media is heterogeneous and promotes many different options. To extend the modelling power to more real cases, however, the external effects can be divided in multiple sources of information, i.e. $I^*=[\vec I^1, \vec I^2,\ldots, \vec I^M]$. In reality, there can be many such information sources, which can be aggregated based on the choice they promote more. For instance, in the case of telephone companies, all magazines advertising one company can be combined into one information source. Consequently, after aggregation, we can consider $K$ sources of information, with $\vec I^i$ promoting more the $i$th option. Each $\vec I^i$ can be in turn modulated to account for extreme and mild information. For the analysis presented here, we consider 
\begin{equation}
I^*=
  \left(\begin{array}{c}
         \vec{I}^1\\
	 \vdots\\
         \vec{I}^K\\
        \end{array} 
  \right)=
  \left(\begin{array}{cccc}
  a&b&\cdots&b\\
  b&a&&\vdots\\
  \vdots&&\ddots&b\\
  b&\cdots&b&a
  \end{array}\right)
\end{equation}  
where each raw is an information source and $b=\frac{1-a}{K-1}$, with  $a\geq b$. Hence, information $\vec I^i$ promotes the adoption of choice $i$ with probability $a$ and the other choices with equal probabilities lower than $a$. If $a$ is large (close to $1$) then each information source is extreme, while when $a$ is close to $b$, they are mild. This choice of $I^*$ assumes all options are equally promoted. This approach can, however, be easily extended to different levels of promotion of the various options.

We consider the same dynamics as in the model with a single opinions source, $p_I$ being now the probability of interacting with one of the information sources. Each individual, at each time step, chooses between the different information sources accordingly to its current opinions: each position $i$ in its opinion vector ($\vec{x}^{\,i}$) is also the probability to interact with the corresponding $\vec I^i$. This signifies that if an agent prefers choice $i$, it will also prefer to interact with sources of information promoting the same choice. For example, by borrowing an analogy from politics, right-wing voters read right-wing oriented newspapers more than left-wing ones. A very extreme agent will interact with other information sources very little, while a moderate agent will interact with more information sources at different times.

\section{Results}
This section presents numerical results for the model introduced. Simulations have been performed with a population of 2000 agents, and 10 different instances for each parameter value. Simulations have been ended when both the number of clusters in the population, and the overlap with the information became stable. A detailed analysis of the effect of $p_I$ for the model with one information source has been presented in \cite{sirbu2012}. Here, we first revisit the study of the effect of the initial condition for $p_I\in\{0,0.01,0.5\}$, with additional insights, then we compare the effect of multiple information sources with previous results.

\subsection{Initial cohesion and number of choices}

 \begin{figure}[th]
\centerline{\psfig{file=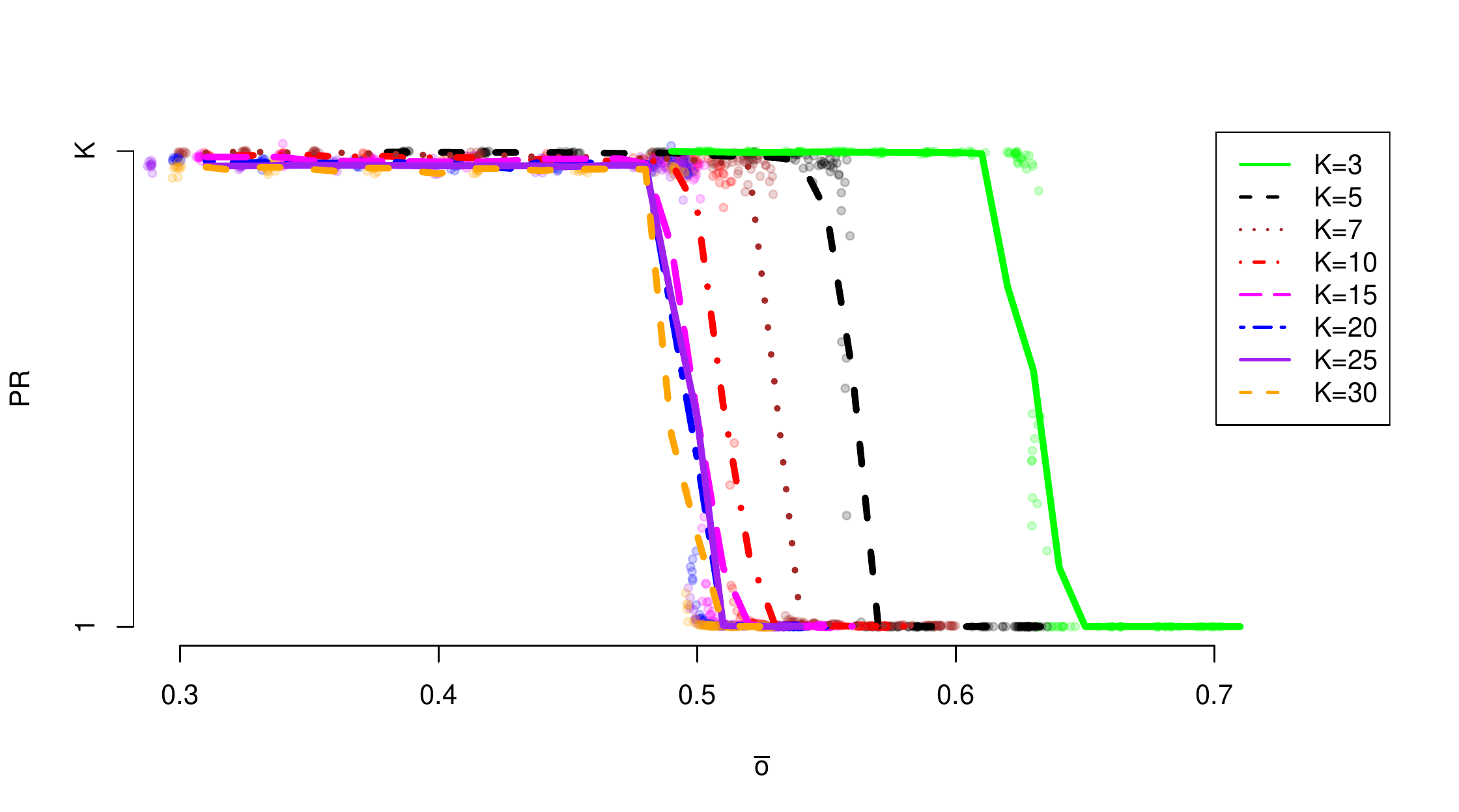,width=12cm}}
\vspace*{8pt}
\caption{$PR$ for different initial cohesion $\bar{o}$ and number of choices $K$ ($N=2000$). The lines represent average values obtained by binning over $\bar{o}$ while points are individual simulation runs. The image shows a transition between $K$ and one cluster as initial cohesion increases. The transition point was previously shown \cite{sirbu2012} not to depend much on the size of the system ($N$). }\label{coh}
\end{figure}

\begin{figure}[th]
\centerline{\psfig{file=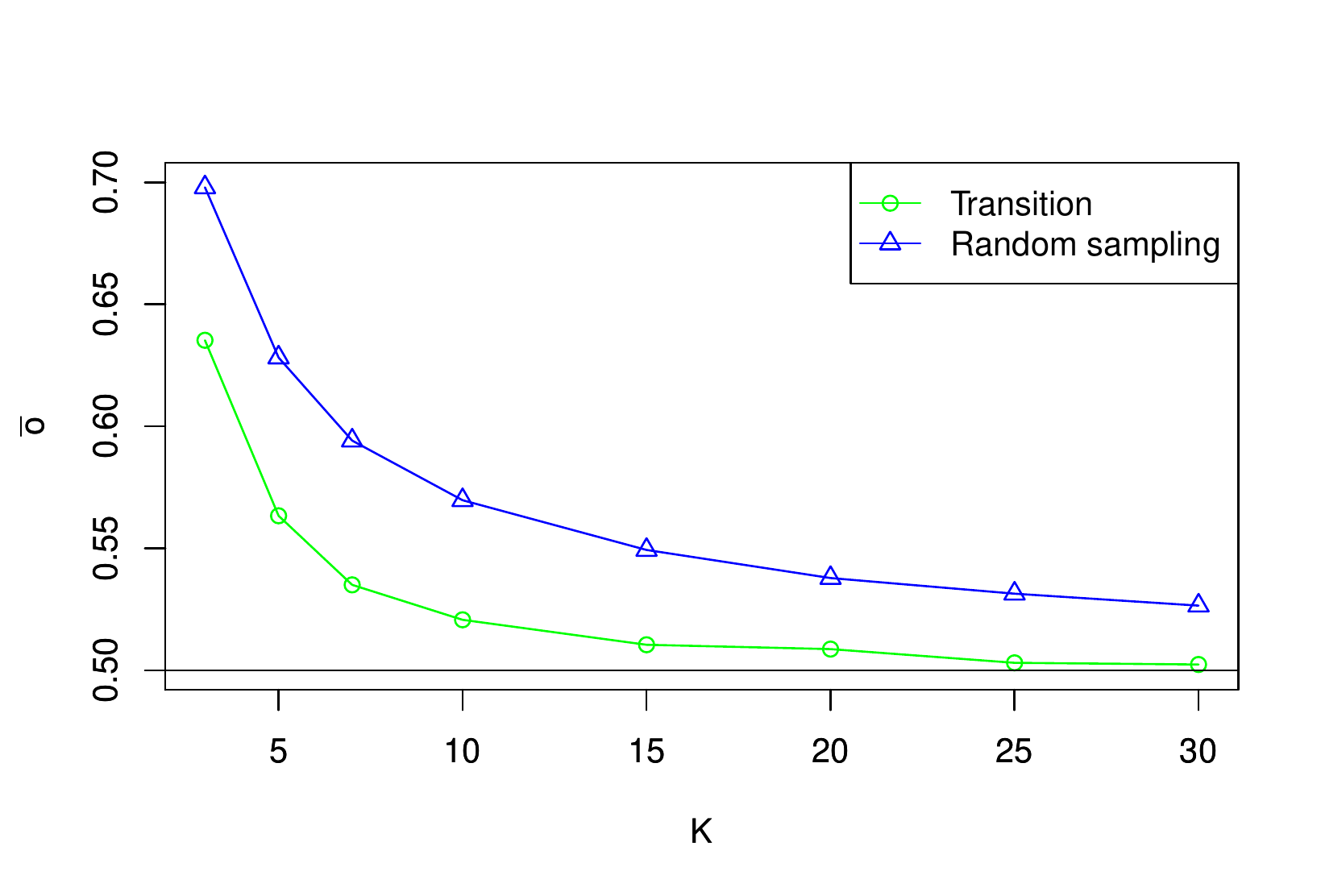,width=12cm}}
\vspace*{8pt}
\caption{Transition point (see Figure~\ref{coh}) and $\bar{o}$ for a random uniform sampling of the simplex, as a function of the number of opinions $K$. The random situation results always in a unique cluster.}\label{cohVsK}
\end{figure}

Firstly, the effect of the cohesion of the initial population in the model with disagreement will be discussed, when no information is present. Initial cohesion is measured as the average pairwise overlap in the starting population:
\begin{equation} 
	\bar{o}=\frac{2\sum_{i,j}o^{ij}}{N(N-1)}.
\end{equation}
This can have a big impact on the dynamics, since it represents the average probability that a randomly chosen pair of individuals will agree. To generate populations with different cohesion, the agent entropy (as defined in Information Theory) is computed:
\begin{equation} 
S_{\mathrm{Ind}}=-\sum_{i=1}^{K}{p_i  \log_2(p_i)}.
\label{entropy}
\end{equation}
\noindent 
We used a random sampling \cite{devroye1986} that yields a uniform density of agents over the simplex. This will produce relatively high cohesion in the population, and depends on the number of choices. To decrease cohesion and obtain a more segregated initial population, individuals with $S$ over a certain threshold $S^*$ are filtered out.

Of interest is the number of clusters obtained in the population for different values of $\bar{o}$, at the end of the opinions formation process (when clusters become stable). Groups are obtained by complete linkage hierarchical clustering \cite{manning1999} with a threshold of 0.8, which guarantees that agent pairs in one cluster have overlap larger than 0.8. The participation rate is used to assess the clustering of the population:
\begin{equation}
PR=\frac{\left(\sum_{i=1}^{C}{c_i}\right)^2}{\sum_{i=1}^{C}{c_i^2}}.
\end{equation}
\noindent This takes into account not only the number of clusters, but also their relative size. For instance, in a population with two clusters, $PR=2$ only if the two clusters are equal in size, and would be very close to 1 if one of the clusters is much larger than the other. Hence, this gives the effective partitioning of the population.

Numerical simulations for a population of 2000 agents, different initial cohesion $\bar{o}$ and various $K$ have been performed, with $PR$ values shown in Figure \ref{coh}. This shows that for low $\bar{o}$ the population is divided into $K$ clusters, while for high $\bar{o}$ one cluster forms. An abrupt transition between the two possibilities appears around a specific value of $\bar{o}$. This transition point depends on the number of opinion choices. Figure \ref{cohVsK} shows how this point decreases as $K$ increases, indicating that consensus is facilitated by large $K$. The figure also shows the average overlap for a random uniform sampling of the simplex. The transition point approaches 0.5 as $K$ increases. At the same time, the average pairwise overlap of a randomly generated population can be easily demonstrated to approach the asymptotic value $0.5$ for infinite $K$. However, for finite $K$, $\bar{o}$ in a randomly generated population remains slightly higher than the transition point, indicating that a population generated by a random sampling from a $K-1$ simplex, even for quite large $K$, will tend ultimately to form a unique cluster.

\subsection{One external information source} 

To study the interplay between the initial condition and the external information, we have performed simulations with different initial cohesion $\bar{o}$ and different information types, from very extreme to very mild. Results are presented for $K=5$. Information extremism can be quantified by computing its entropy $S_I$:
\begin{equation} 
S_{I}=-\sum_{i=1}^{K}{I_i  \log_2(I_i)}.
\label{entropyInfo}
\end{equation}
This takes high values for mild information and low for extreme. All figures display a normalised value for the entropy (${S_I}/{\log_2{K}}$). Hence, for $[1,0,0,0,0]$ a value of 0 indicates maximum extremism, while for $[0.2,0.2,0.2,0.2,0.2]$ a value of 1 indicates maximum mildness. Results are studied in terms of cluster number ($PR$) and information success. The latter was computed as the average overlap of the information with the agents in the population: 
\begin{equation} 
\overline{IO}=\frac{1}{N}\sum_{i=1}^{N}{o^{Ii}} 
\end{equation}
with
\begin{equation} 
 o^{Ii}=
    \frac{\vec{I}\cdot\vec{x}^{\,i}}{|\vec{I}| |\vec{x}^{\,i}|} =
    \frac{\sum_{k=1}^{K}{I_k p_k^i}}{\sqrt{\sum_{k=1}^{K}{{I}^2} \sum_{k=1}^{K}{(p_k^i)^2}}}.
\end{equation}
\begin{figure}[th]
\centerline{\psfig{file=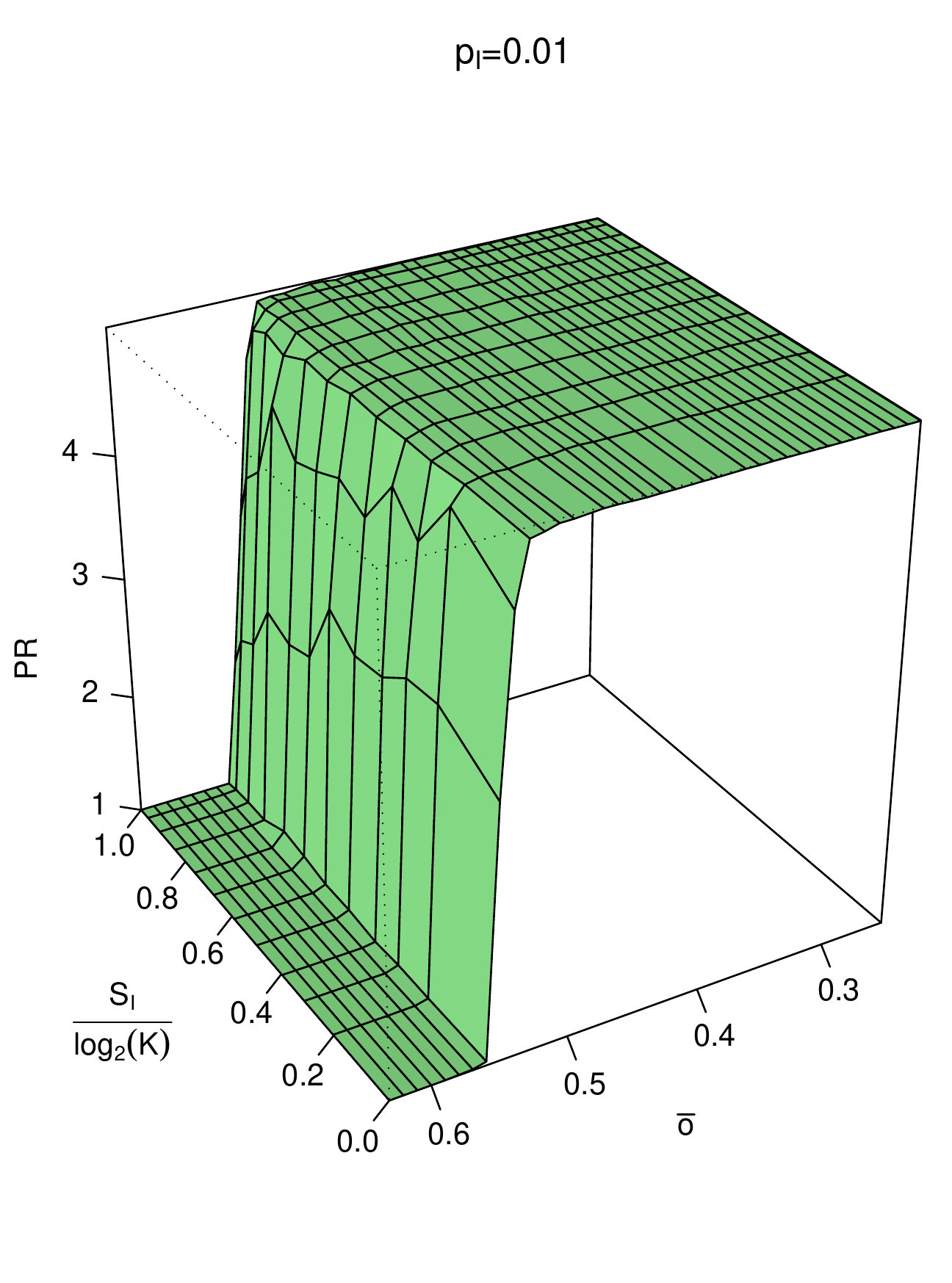,width=6cm}\psfig{file=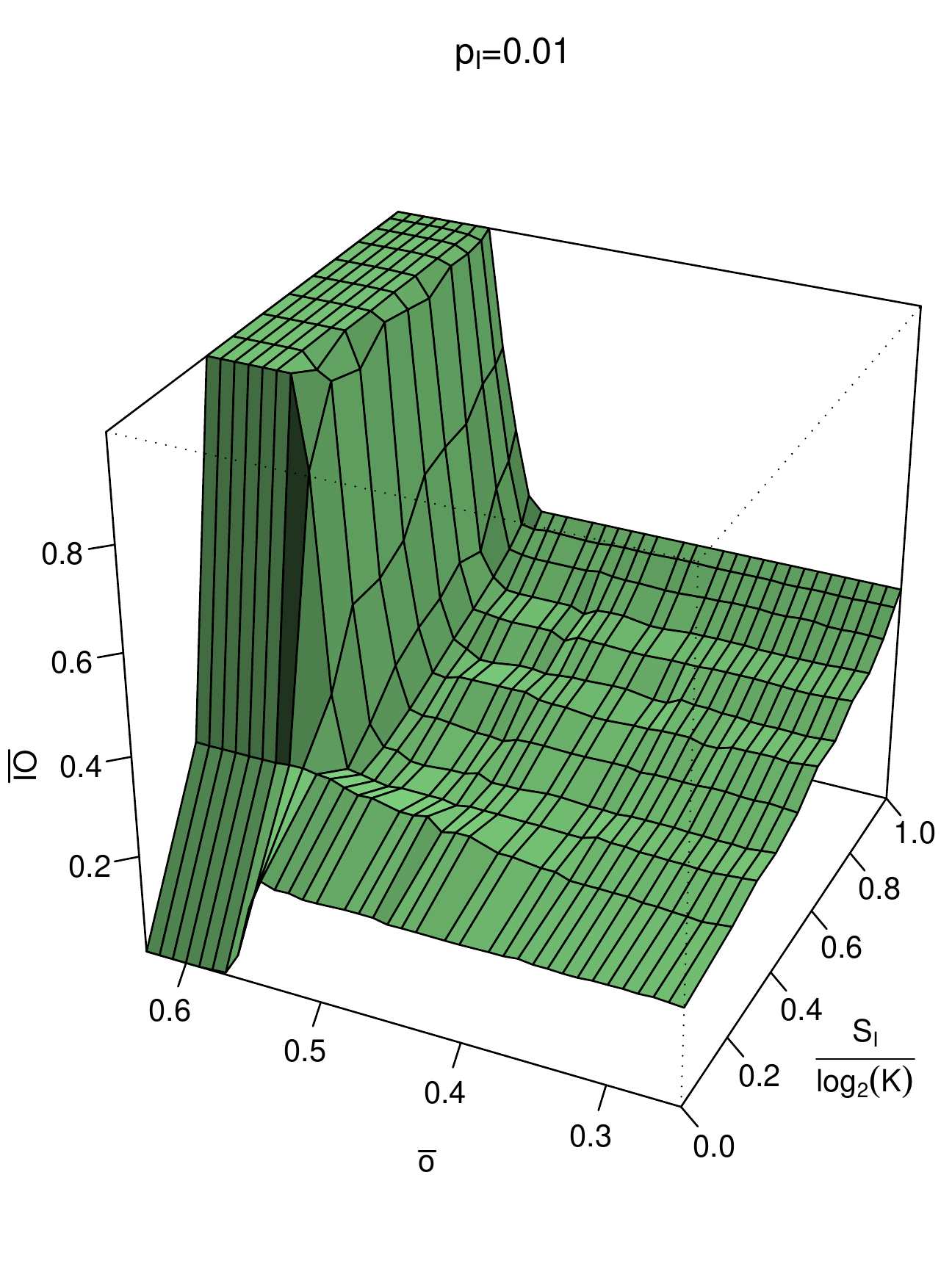,width=6cm}}
\caption{One information source: number of clusters ($PR$) and average overlap of the information with the agents in the population ($\overline{IO}$). Results of the simulations after reaching the convergence are reported as a function of  different initial average pairwise overlap in the population ($\bar{o}$) and different information extremism levels,
as measured by the normalized information entropy (${S_I}/{\log_2{K}}$). We consider here low exposure to information ($p_I=0.01$).}\label{oneInfopi0.01}
\end{figure}

Figure \ref{oneInfopi0.01} shows number of clusters ($PR$) and average information overlap ($\overline{IO}$) for slow exposure to information, $p_I=0.01$. A low $p_I$ does not affect the number of clusters in the population, with the initial cohesion ($\bar{o}$) driving cluster formation. So, five clusters are obtained for segregated initial populations, and one for compact populations. Information success ($\overline{IO}$), however, changes as $\vec{I}$ becomes milder. In the case of low initial cohesion, when the population splits into five clusters, $\overline{IO}$ values are low and increasing slightly as ${S_I}/{\log_2{K}}$ becomes 1 (i.e. information becomes milder). In this case, the clusters are of the form [1,0,0..], so very extreme, and the low $p_I$ limits the power of the information to attract individuals. Conversely, when the initial condition are such that the population would form one cluster even in the absence of external information, two different scenarios emerge depending on the 
information extremism value ($S_I$):
for mild information (${S_I}/{\log_2{K}}\geq 0.4$) the population will eventually be fully attracted by the information, i.e. $\overline{IO}=1$, while when ${S_I}/{\log_2{K}}$ is close to 0, the entire population moves away from the external signal ($\overline{IO}=0$), since it is too extreme.

\begin{figure}[th]
\centerline{\psfig{file=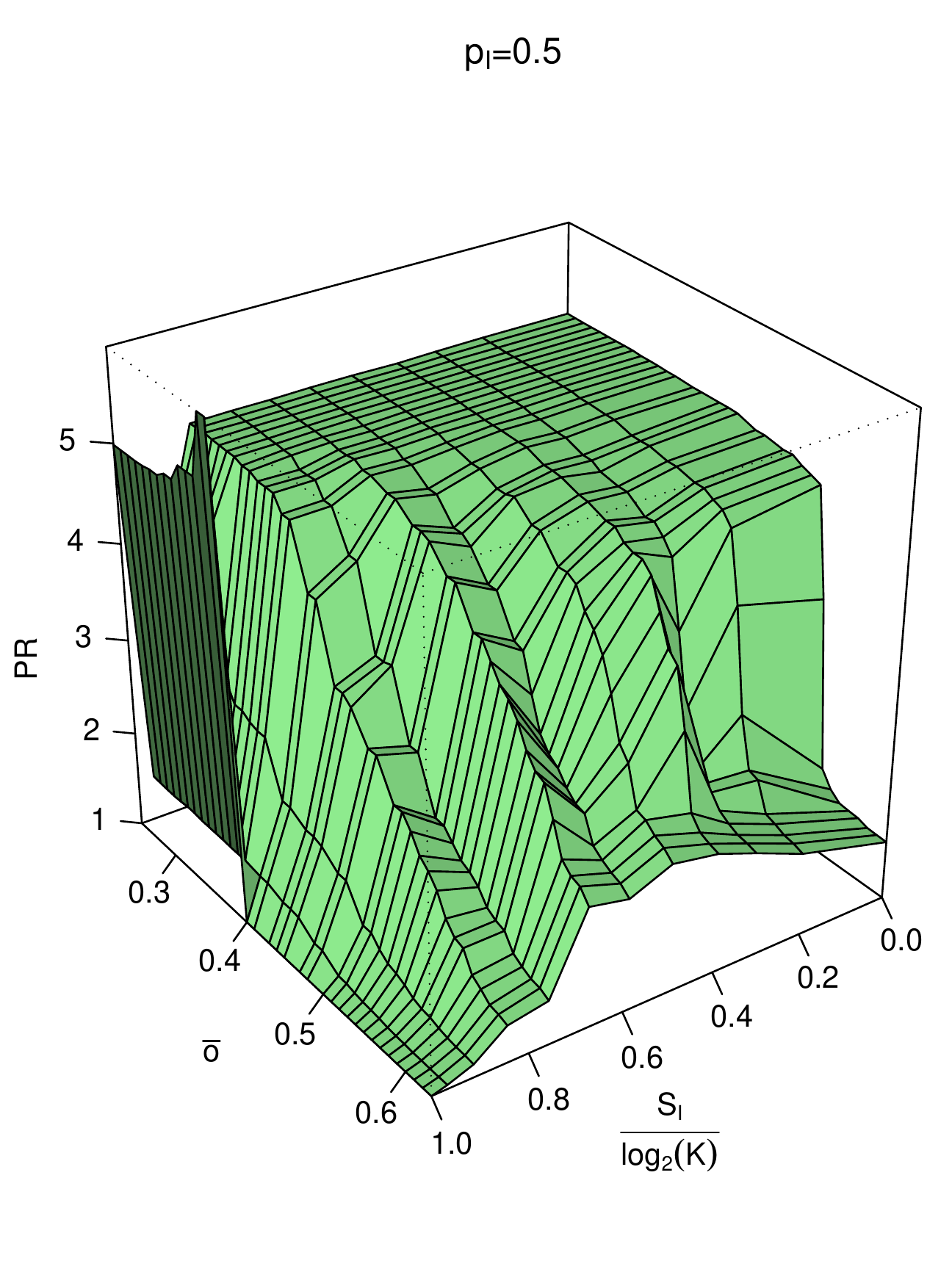,width=6cm}
\psfig{file=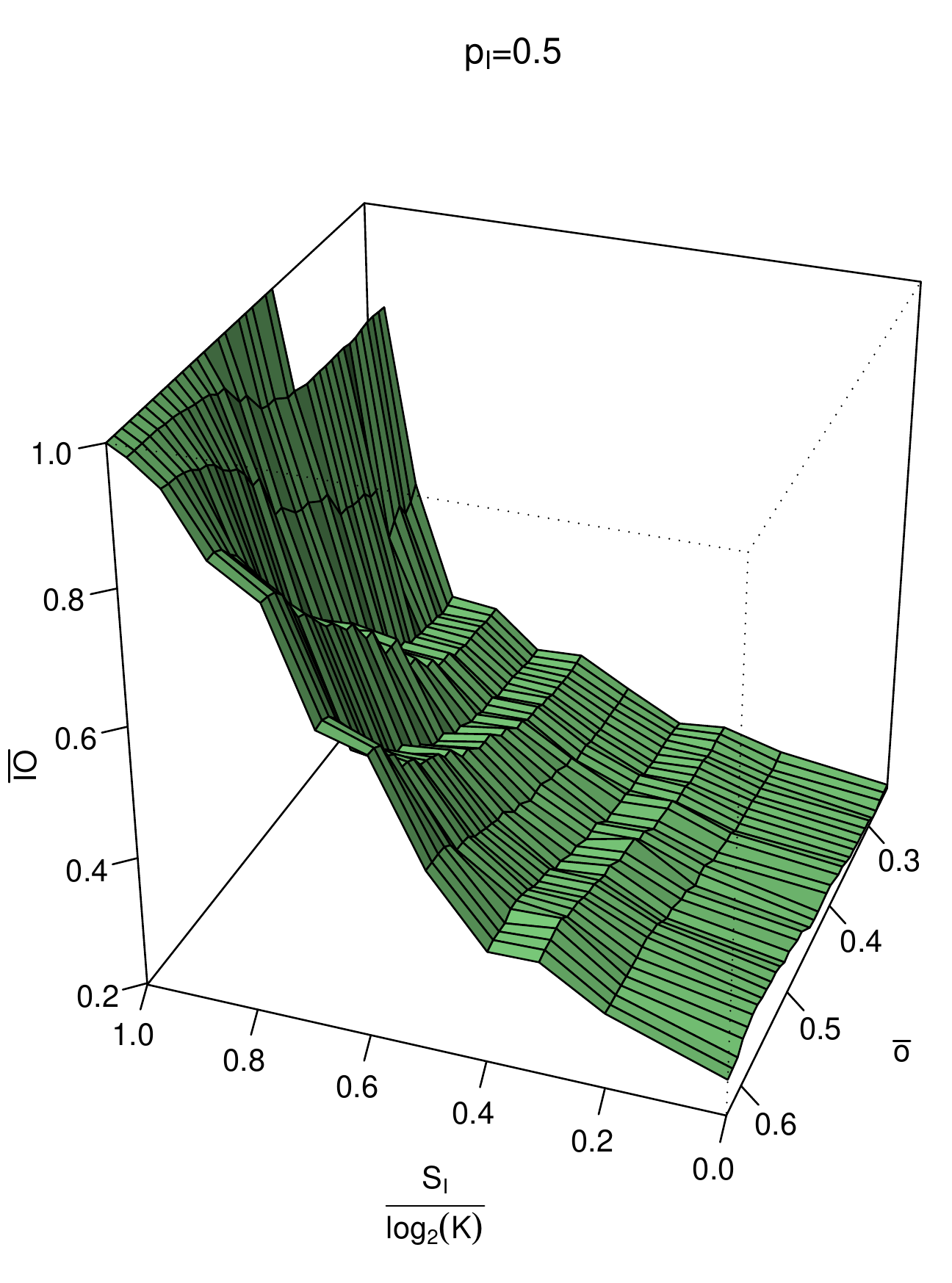,width=6cm}}
\caption{Number of clusters ($PR$) and information success ($\overline{IO}$) for different initial conditions and information extremism levels (${S_I}/{\log_2{K}}$), in the case of high exposure to information ($p_I=0.5$).}\label{oneInfopi0.5}
\end{figure}

Increasing exposure to information ($p_I=0.5$) changes system behaviour. Figure \ref{oneInfopi0.5} shows $PR$ and $\overline{IO}$ for this case. As information becomes milder (${S_I}/{\log_2{K}}$ increases), the number of clusters decreases in the population, indicating that mild information favours consensus. This happens even when the initial population is segregated (low initial cohesion $\bar{o}$), so when the natural trend in the population would be to form five clusters. On the other hand, extreme information favours segregation, with more than one cluster obtained even when initial cohesion is very large. At the same time, the information success ($\overline{IO}$) increases as information becomes milder, with a maximum value obtained for very mild information and large initial cohesion. An interesting effect can be observed when information has maximum entropy, i.e. it promotes equally all choices, hence it does not hold an informative message. In this case, the initial cohesion $\bar{o}$ drives again the number of clusters, with five clusters for low $\bar{o}$ and one for large $\bar{o}$. Around the transition, however, a further cluster appears, most probably due to the indecision of the external information. This is formed by undecided individuals ($[0.2,0.2,0.2,0.2,0.2]$) who coexist with five clusters of extremists. This situation (more than 5 clusters) does not appear when the external input carries some useful information.

It is important to note that full information success ($\overline{IO}=1$) is obtained when $p_I=0.5$ only for the mildest possible information (${S_I}/{\log_2{K}}=0$). At the same time, for low information exposure (low $p_I$), full agreement can be obtained, if the population is initially compact, even for more extreme information types. This indicates that aggressive media campaigns are not beneficial, especially if they contain extreme messages. Also, an analysis of the target group is necessary to enhance information success.

\begin{figure}[th]
\centerline{\psfig{file=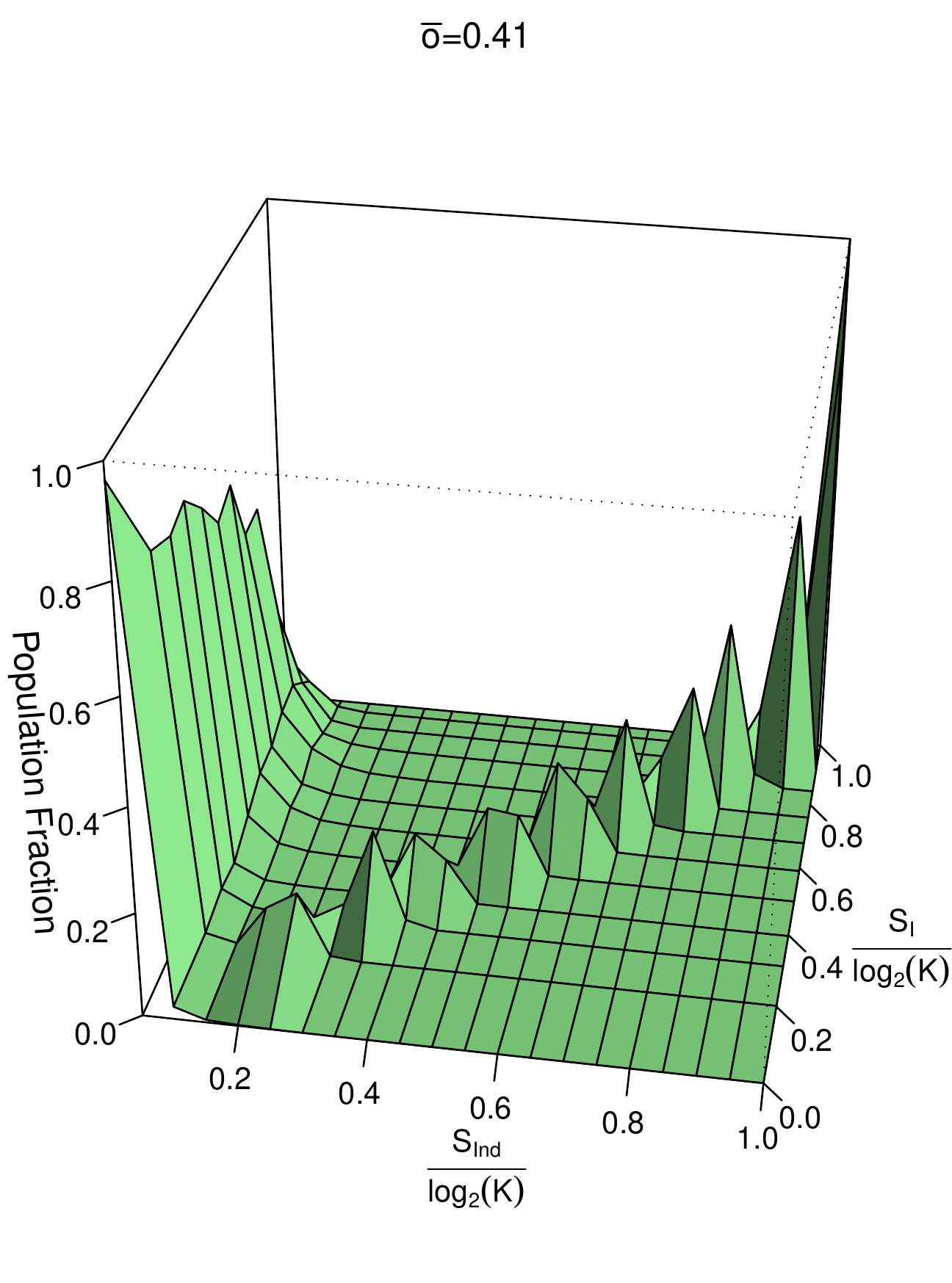,width=6cm}
\psfig{file=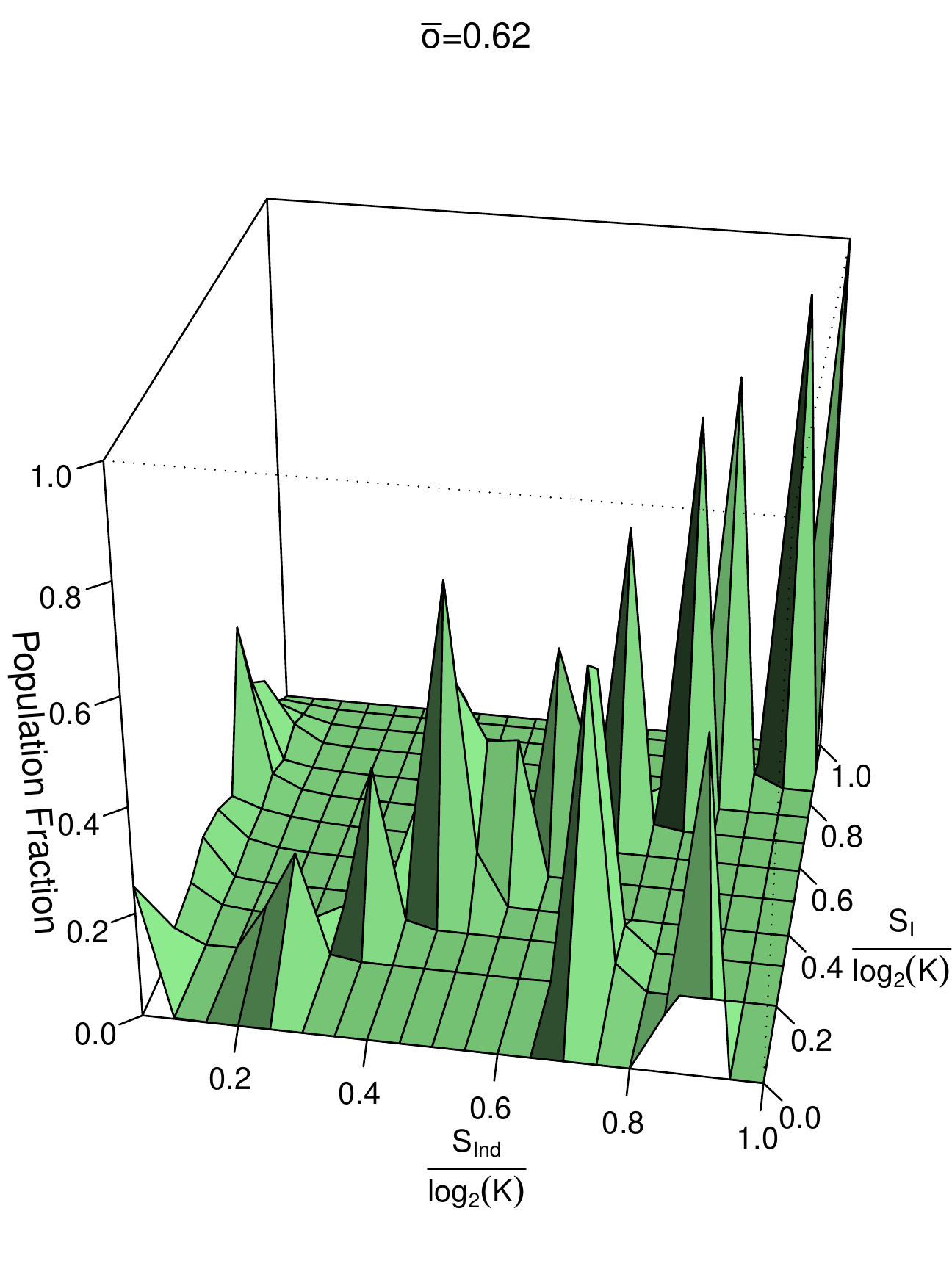,width=6cm}}
\caption{Distribution of individual indecision (normalised individual entropy ${S_\mathrm{Ind}}/{\log_2{K}}$) for populations obtained with different external information types (various information entropy levels ${S_I}/{\log_2{K}}$)  for two initial conditions: $\bar{o}\in\{0.41,0.62\}$;
($p_I=0.5$).}\label{indEntropyOneInfo}
\end{figure}

Although the average agreement to the information and the number of clusters are important when modelling opinion dynamics, the structure of the individual opinions is meaningful as well. Figure \ref{indEntropyOneInfo} shows the state of indecision of individuals obtained for different information values and two initial conditions (with small and large cohesion), when $p_I=0.5$. Indecision is measured through the agent's entropy $S_\mathrm{Ind}$, and here normalised values are displayed - ${S_\mathrm{Ind}}/{\log_2{K}}$ (1 for maximum indecision, 0 for a very decided opinion). The figure shows histograms of the indecision levels of all individuals in the population. 

\begin{itemize}

\item[$\bullet$] For a segregated initial population ( $\bar{o}=0.41$ - when multiple clusters exist in the population - left panel), the figure shows that some individuals follow the level of indecision of the information (non zero value of ${S_\mathrm{Ind}}/{\log_2{K}}$), while the rest form extremist groups (peaks at ${S_\mathrm{Ind}}/{\log_2{K}}=0$) promoting only one option. The extremist groups decrease as the information becomes milder (${S_{I}}/{\log_2{K}}$ increases) and disappear only when information becomes $[0.2,0.2,0.2,0.2,0.2]$. Hence, in this condition, full agreement to the information ($\overline{IO}$) is only obtained by an external signal carrying no useful information. For other information values, the size of the cluster adhering with the information, and the level of indecision, increase as information become milder.

\item[$\bullet$] In the case where the initial cohesion is large ($\bar{o}=0.62$ - the number of clusters in the population is smaller -right panel), population clusters form again around the level of indecision of the information, however the extremist individuals are more rare, and additional undecided individuals appear, especially when information is extreme (${S_{I}}/{\log_2{K}}$ is low). For instance, for the case $I=[1,0,0,0,0]$ (${S_{I}}/{\log_2{K}}=0$), a small part of the population adheres with the information, while the rest oppose the information, remaining undecided on the other positions (values around $[0,0.25,0.25,0.25,0.25]$).

\end{itemize}

All in all, using one external source of information allows for important observations to be made regarding the effect of the initial condition and the success of extreme messages. However, the structure of the population does not apply for all real situations. Multiple clusters appear but many contain very extreme agents. At the same time, full agreement in the population is obtained mostly when individuals are undecided. The only exception is for low information exposure and compact initial conditions. In some real situations, individuals rarely hold a very decided opinion, as more options can have some value even if one is favourite. At the same time, extremely mild opinions represent undecided individuals, and general states of indecision in a population are rare. To be able to model these situations as well, multiple information sources can be introduced, and an analysis will be presented in the following section.

\subsection{Multiple information sources}
A similar analysis of the number of clusters and information success when the initial cohesion of the population and information extremism changes is performed here, again for $K=5$ and $N=2000$. To obtain different information extremism levels, $a$ (refer to section~\ref{sec:MI}) takes different values in $[1,0.2]$, with the five information vectors having the same normalised entropy (${S_I}/{\log_2{K}}$) ranging from 0 to 1. The number of clusters is assessed using the same $PR$ as above. For information success, however, the average information overlap is not relevant enough, since now there are five different information vectors. One of the measures of interest, however, is whether some information source is more successful than others. To test this, we compute the average overlap with each of the five information vectors, similar to the previous section, then we compute the entropy of the obtained array (after normalisation). This entropy of information overlap ($S_{IO}$) can distinguish between situations where all information vectors are equally successful and where some have more success than others.

\begin{figure}[th]
\centerline{\psfig{file=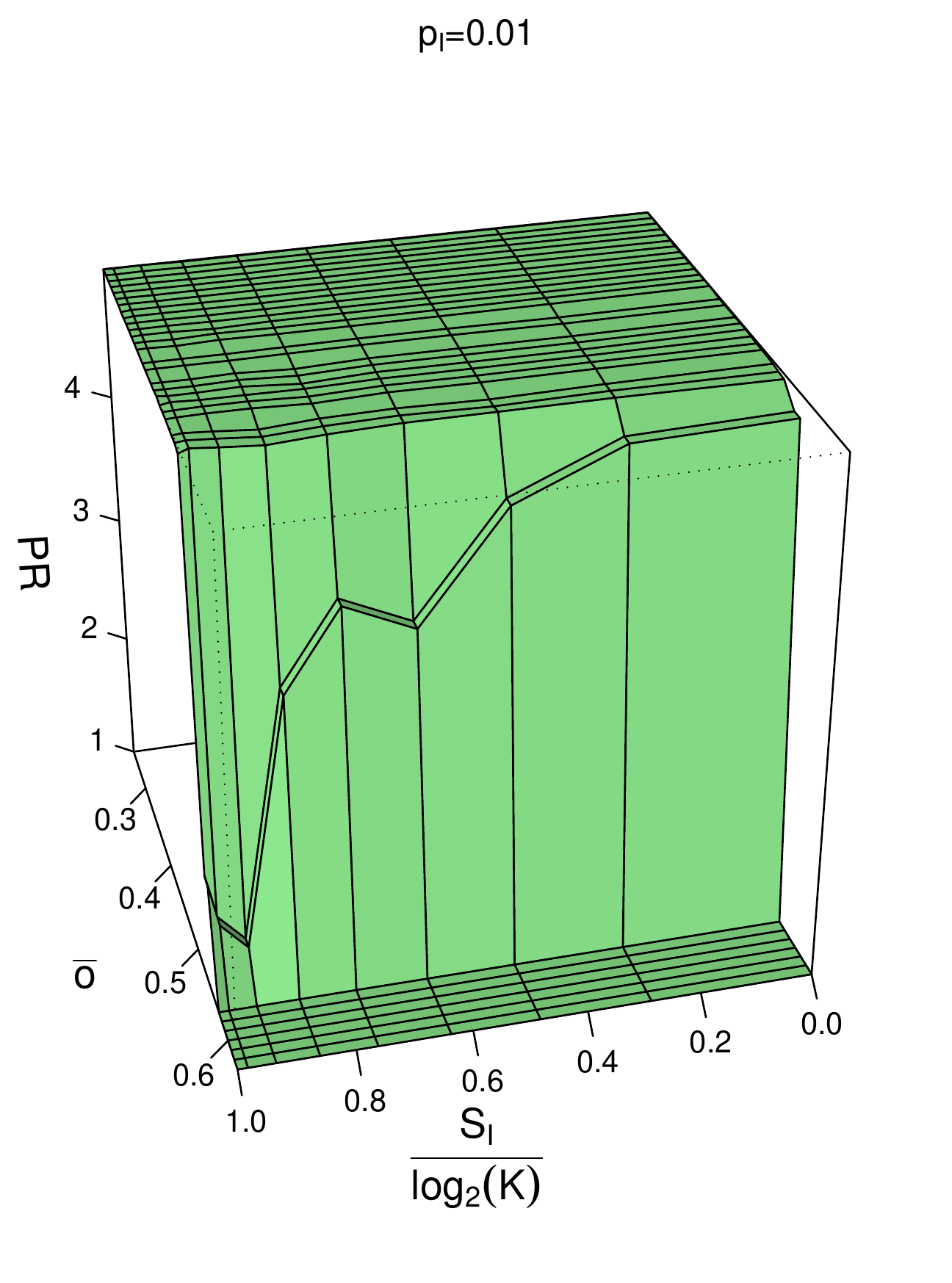,width=6cm}
\psfig{file=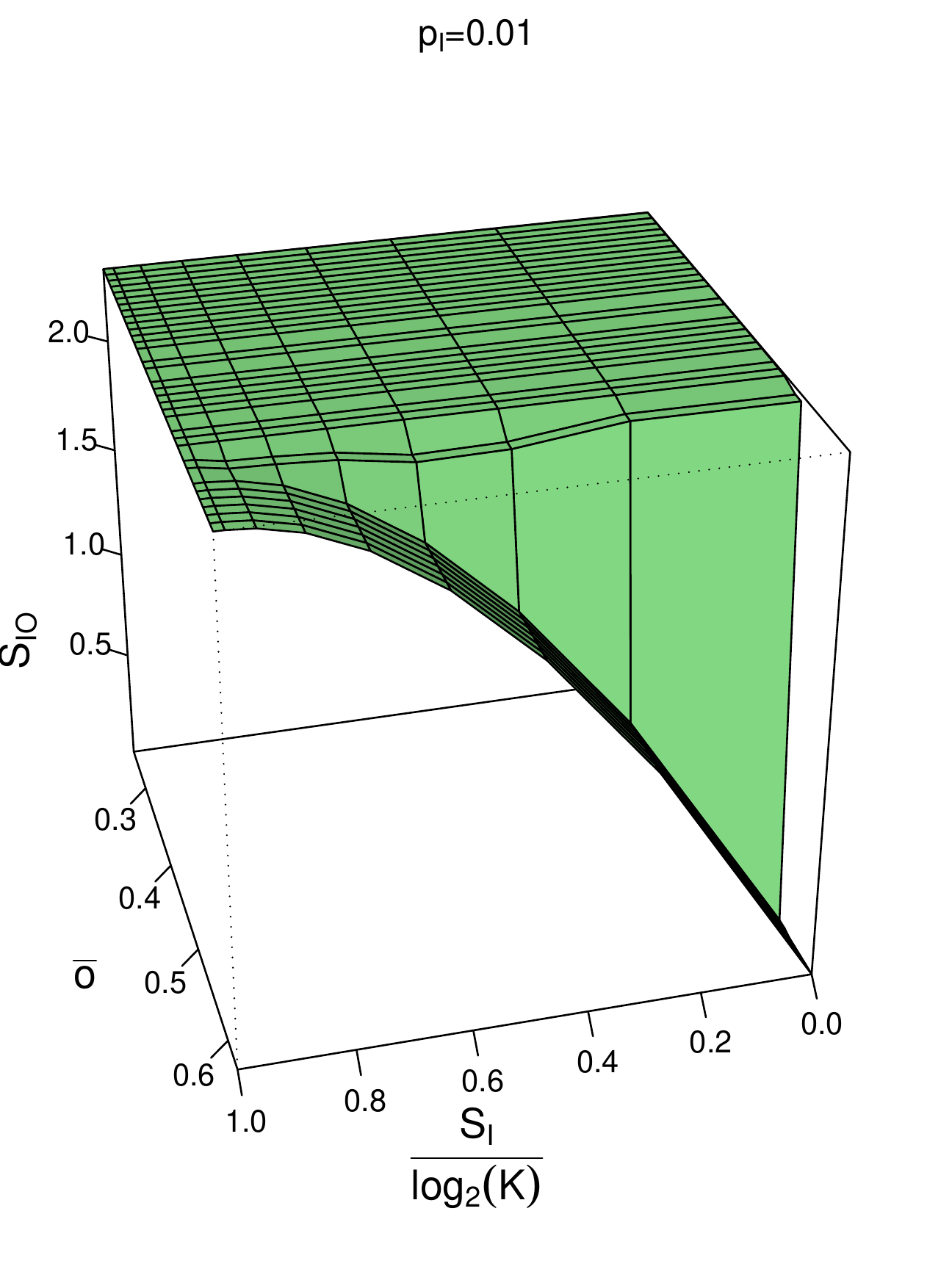,width=6cm}}
\caption{Multiple information sources: number of clusters ($PR$) and entropy of information overlap ($S_{IO}$) for different initial conditions ($\bar{o}$) and information extremism levels (${S_I}/{\log_2{K}}$ - a function of parameter $a$), in the case of low exposure to information ($p_I=0.01$).}\label{moreInfopi0.01}
\end{figure}
 
Figure \ref{moreInfopi0.01} shows $PR$ and $S_{IO}$ values for low exposure to information ($p_I=0.01$). Similar to the case of one information source, the number of clusters depends only on the initial condition, with five when the initial population is segregated (low $\bar{o}$) and one otherwise: 

\begin{itemize} 

\item[$\bullet$] When five clusters appear, $S_{IO}$ values (close to maximum possible value of $\log_2(K)$) indicate that the population is equally divided among the five information sources.

\item[$\bullet$] When one cluster is present, however, $S_{IO}$ decreases, indicating that the cluster forms around one of the possible information, as long as the information is not extremely mild (${S_I}/{\log_2{K}} < 1$, corresponding to $a \ge 0.4$, which means one opinion choice has probability over 40\% and the others under 15\%). When the information is mild (in our simulations, for $a\in\{0.3,0.2\}$), a general state of indecision appears where none of the information sources fully attracts individuals (high value of $S_{IO}$).  \end{itemize}

\begin{figure}[th] \centerline{\psfig{file=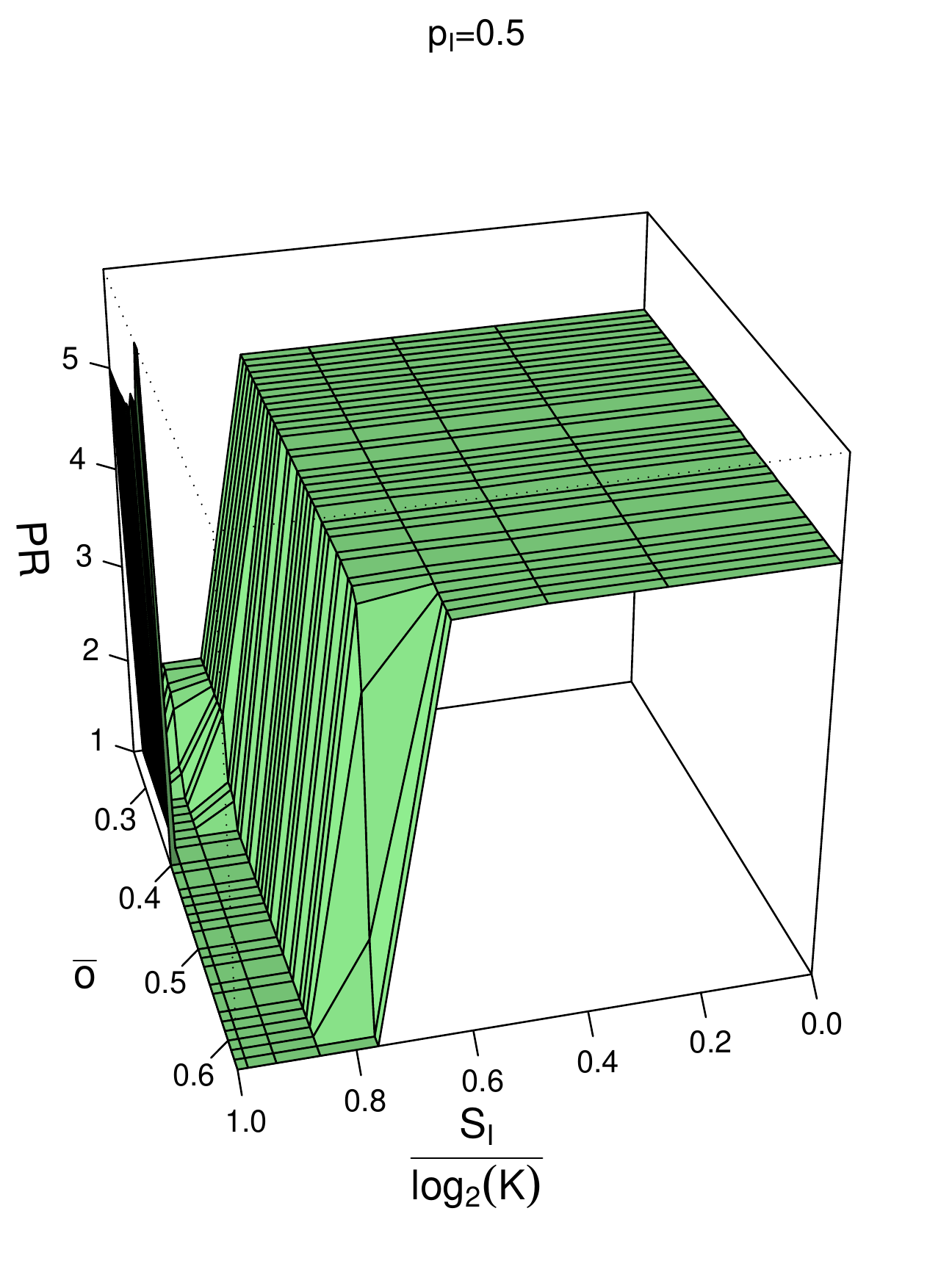,width=6cm} \psfig{file=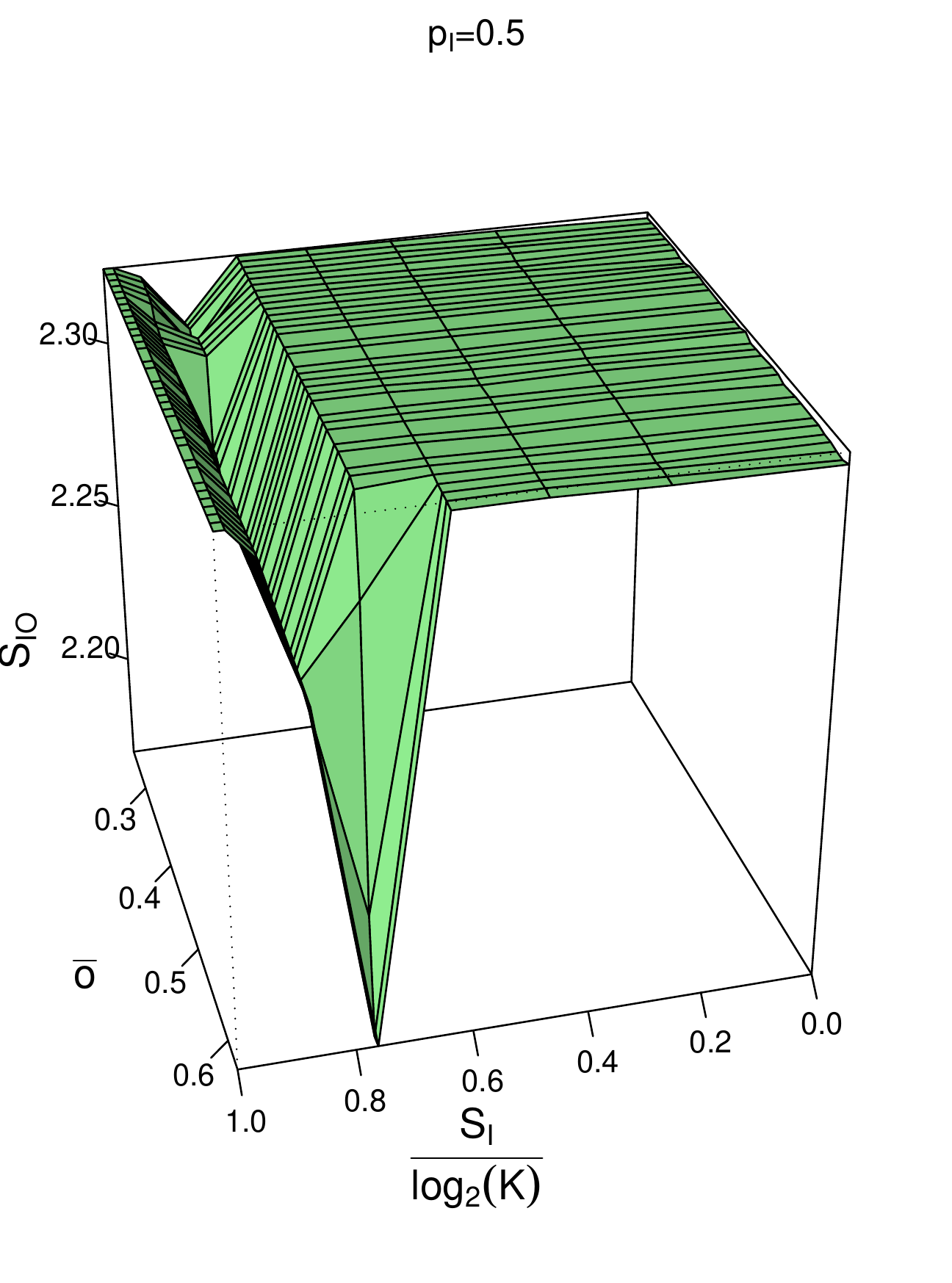,width=6cm}} \caption{Multiple information sources: number of clusters ($PR$) and entropy of information overlap ($S_{IO}$) for different initial conditions ($\bar{o}$) and information extremism levels (${S_I}/{\log_2{K}}$ - a function of parameter $a$), in the case of high exposure to information ($p_I=0.5$).}\label{moreInfopi0.5}
\end{figure}

For $p_I=0.5$, Figure \ref{moreInfopi0.5} shows that introducing $K$ information sources makes the system more sensitive to the external information. Specifically: 
\begin{itemize} 
\item[$\bullet$] The segregation effect of extreme information is amplified, with five clusters obtained even for large cohesion in the initial population ($\bar{o}$) when ${S_I}/{\log_2{K}}<0.6$ ($a\geq0.7$). In the case of one information source, extreme information divided the population into two clusters, one agreeing and one disagreeing with it.  
\item[$\bullet$] At the same time, milder information leads to one cluster, as long as the initial population is not extremely segregated ($\bar{o}>0.4$). This cluster adheres to one of the five information sources, as the low $S_{IO}$ values demonstrate, provided the information is not too mild ($a>0.3$, similar to the case $p_I=0.01$).  
\item[$\bullet$] When the initial cohesion is very small ($\bar{o}<0.4$), the transition from five to one cluster as the information vectors become milder (${S_I}/{\log_2{K}}$ increases) is not as abrupt, with a region where the two opposite effects (segregation from initial condition and cohesion from mild information) are similar in size, resulting in an intermediate number of clusters agreeing with more than one information source.  \item[$\bullet$]When $a=0.2$ all five information sources are equal so the same situation as for one information with maximum mildness appears.  
\end{itemize}

\begin{figure}[th] \centerline{\psfig{file=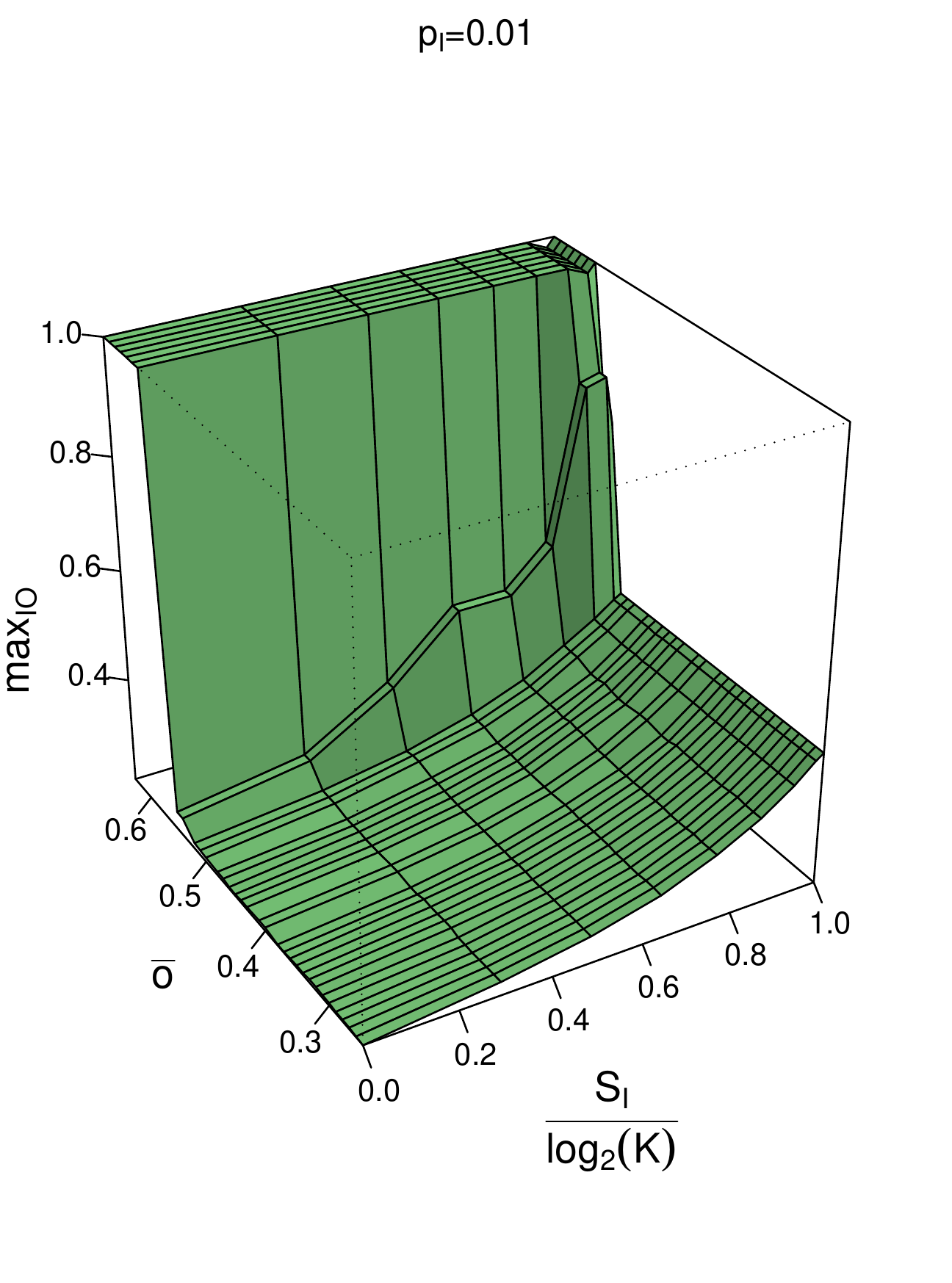,width=6cm} \psfig{file=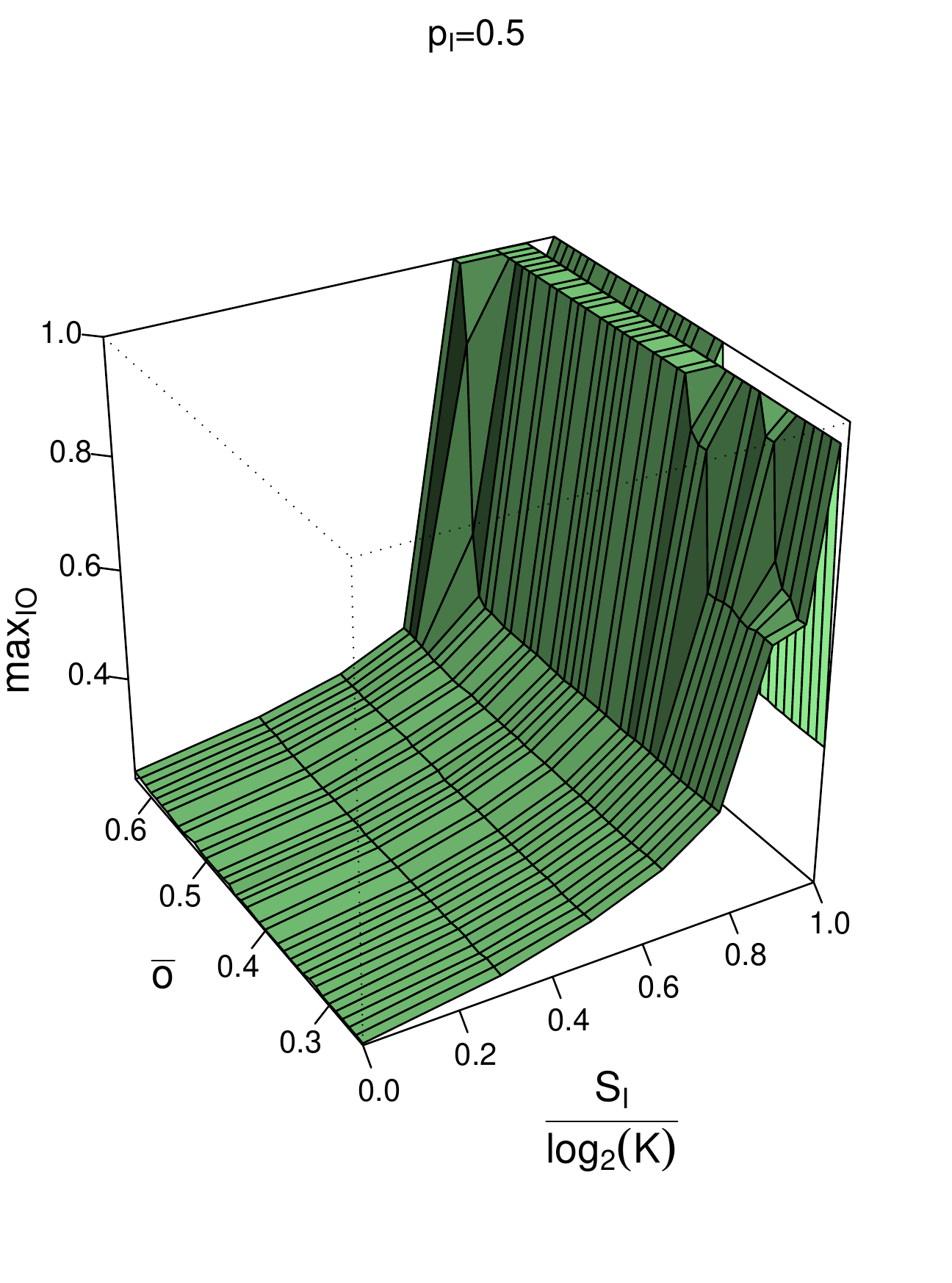,width=6cm}} \caption{Multiple information sources: maximum average information overlap (max$_{IO}$) among the five external information sources vs. the average initial overlap in the population, $\bar{o}$, and ${S_I}/{\log_2{K}}$.}\label{moreInfoMaxO} \end{figure}

For a better picture of the results above, the maximum information overlap (max$_{IO}$) among the five information values is presented, for each parameter set, in Figure \ref{moreInfoMaxO}. If (max$_{IO}=1$), it indicates complete agreement of the population with one of the available information sources, while max$_{IO}<1$ means that no information is fully successful. This plot shows that in all cases from Figures \ref{moreInfopi0.01} and \ref{moreInfopi0.5} when one cluster appears ($PR=1$) and $S_{IO}<\log_2{K}$ (the maximum entropy possible), the maximum information success is 100\% of the population, i.e. the entire population forms one cluster around one of the information sources.

\begin{figure}[th] \centerline{\psfig{file=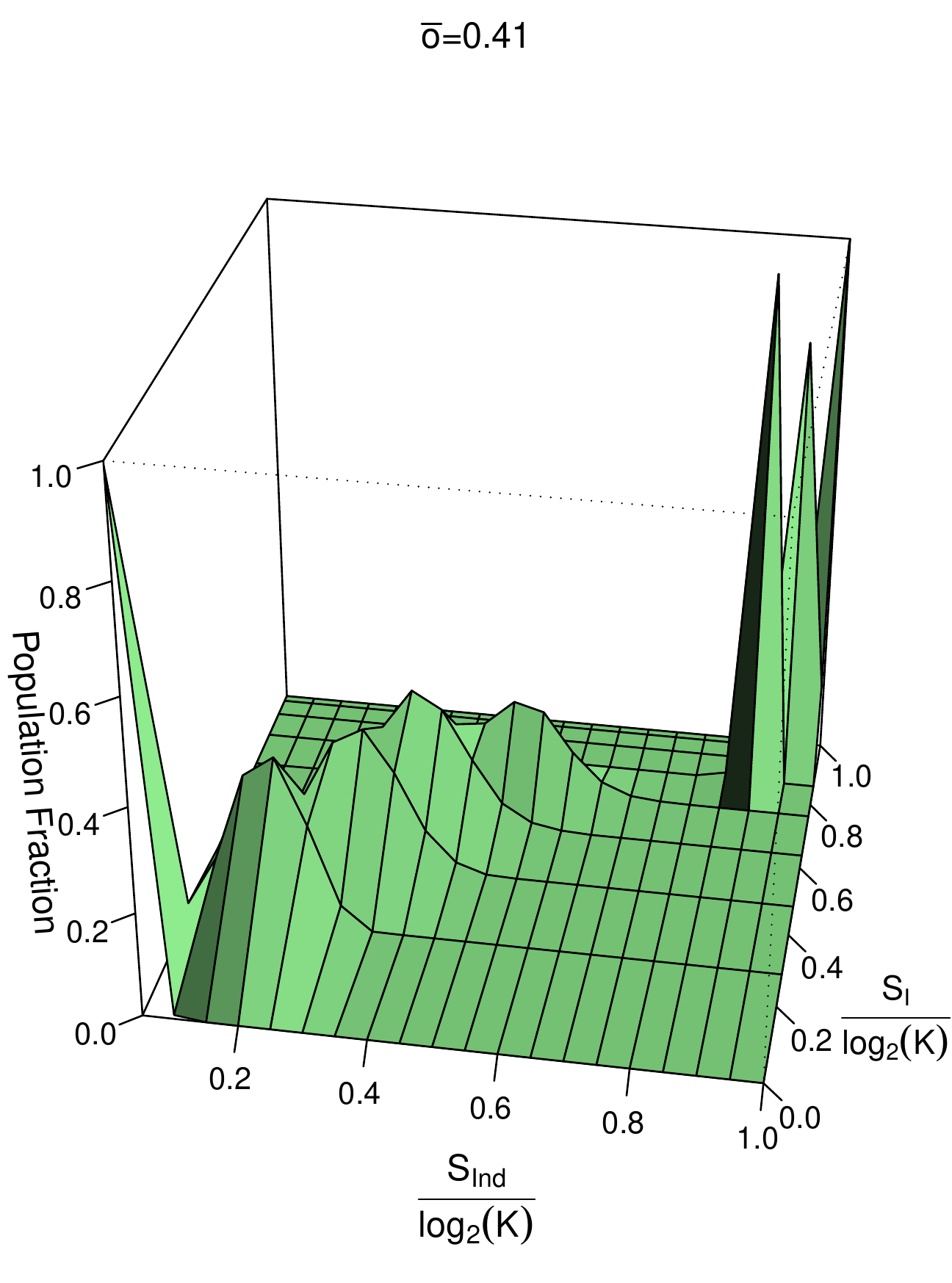,width=6cm} \psfig{file=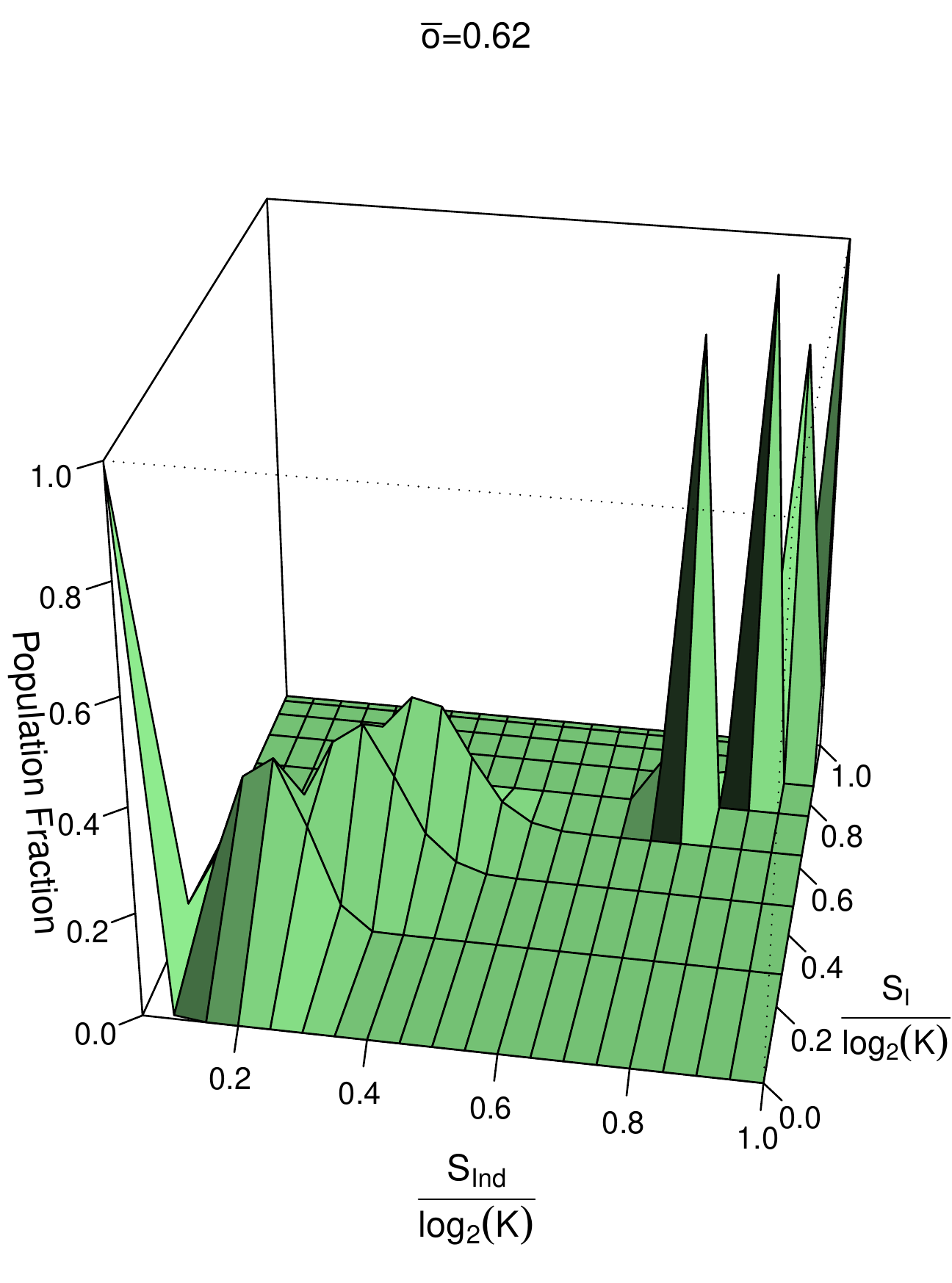,width=6cm}} \caption{Multiple information sources: distribution of individual indecision (normalised individual entropy ${S_\mathrm{Ind}}/{\log_2{K}}$) for populations obtained with different external information types (various information entropy levels ${S_{I}}/{\log_2{K}}$) for two initial conditions: $\bar{o}\in\{0.41,0.62\}$.}\label{indEntropyMoreInfo} \end{figure}

\noindent To analyse the structure of the individual opinions in the population, i.e. their indecision, Figure \ref{indEntropyMoreInfo} shows distribution of normalised agent entropy (${S_\mathrm{Ind}}/{\log_2{K}}$) for two initial conditions and $p_I=0.5$, as the external information sources change from extreme to mild (${S_{I}}/{\log_2{K}}$ increases), similar to the case of one information source.  
\begin{itemize} 
\item[$\bullet$] For the segregated initial condition ($\bar{o}=0.41$, five clusters), the indecision of individuals follows that of the five information sources, and there are no additional extreme clusters forming. Additionally, when information is milder, the full agreement of the population with one of the information is obtained. This happens when information is still meaningful (${S_{I}}/{\log_2{K}} \sim0.85$ i.e. $a=0.5$ meaning one opinion choice has 50\% probability while the others 12.5\%), unlike the situation with one information source, when full agreement appeared only when the external signal was very flat (corresponding to $a=0.2$).  
\item[$\bullet$] For compact initial conditions ($\bar{o}=0.62$) the situation is similar, with either five clusters that are not extreme (unless information itself is extreme) or one cluster that is not too undecided (${S_{I}}/{\log_2{K}}\sim 0.8$ i.e. $a=0.6$).  
\end{itemize}

\begin{figure}[th] \centerline{\psfig{file=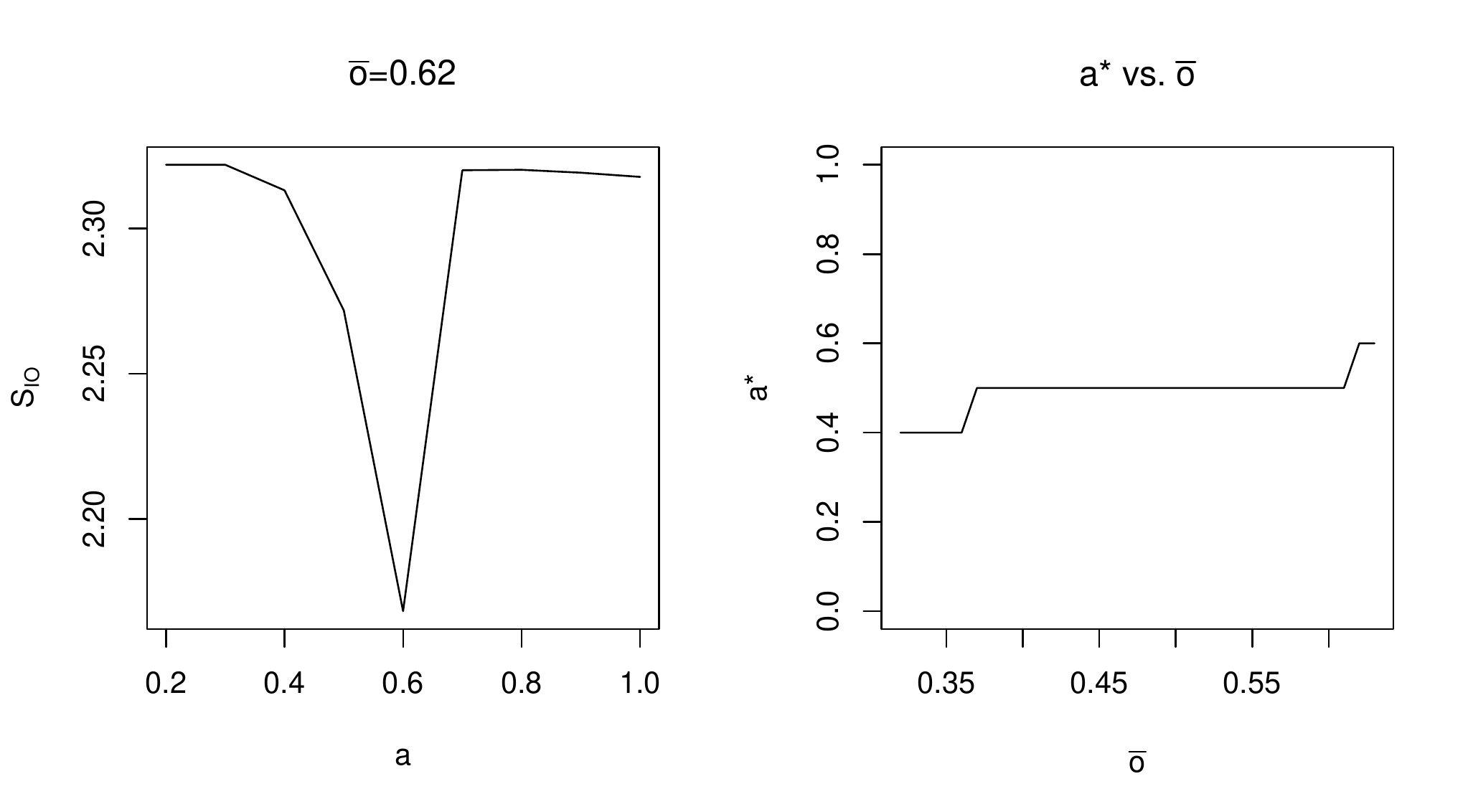,width=11cm}} \caption{Dependence of the agreement with the information ($S_{IO}$) on $a$, and the maximum $a$ for which one cluster is obtained ($a^*$), for different initial conditions ($\bar{o}$).}\label{aOneCluster}
\end{figure}

All in all, results of this section indicate that multiple information sources are crucial to obtain a stable cluster configuration where individuals are not in majority extremists, as opposed to what we have observed for one information source. At the same time, multiple information sources enable full agreement of the population with an external source at lower indecision levels (less mild information), compared to the single information case. The minimum mildness level (maximum $a$) for which the population evolution changes from forming multiple to one cluster depends on the initial condition. Figure \ref{aOneCluster} (left) shows this transition for a compact population ($\bar{o}=0.62$) by plotting the $S_{IO}$ as a function of $a$ (which determines the information entropy or mildness). The transition from five to one cluster happens when $S_{IO}$ falls under the maximum level, at a specific value for $a$, $a^*$. For $\bar{o}=0.62$, $a^*=0.6$ The increase in $a^*$ (hence decrease in information mildness required to obtain one cluster) with the initial cohesion is also displayed in Figure \ref{aOneCluster} (right). For a compact population, a more extreme information can lead to one cluster around one of the four information values, while as the initial cohesion decreases, the information has to become milder for the transition to occur. This indicates again that a careful analysis of the target population is required to maximise information success.

\section{Conclusions}
An analysis of a model for opinion dynamics with disagreement and external information was presented. The initial condition was shown to have a large effect on the population, especially when the external effect was removed. A transition between one and more clusters was observed when initial cohesion in the population decreased. The transition point was shown to decrease with size of the opinion vector, indicating the a larger number of choices facilitates agreement. This has been previously observed for another continuous model of opinion dynamics \cite{Lorenz2007} 

Analysing the model under the effect of an external source of information (which can be modulated from mild to extreme) enabled several observations to be made. Mild information was shown to have a cohesive effect and a larger success in the population, while extreme information produced segregation and had a limited success. Information success was maximised for mild messages and slow information exposure, indicating that aggressive advertising campaigns (both in volume and message type) are not beneficial. Also, the initial condition was very important in driving the results, which means a careful analysis of the target group is necessary when designing a media campaign. 

Using one information source gave important insights into the different effects an external signal can cause. However, the structure of the population was shown not to be suitable for all real situations, since multiple clusters came with extremism while a single cluster in general meant indecision. Although there are real situations when opinion clusters can take mostly extreme values (for instance religious views), in many real-life situations people consider more options valuable, with one being the favourite (e.g. telephone companies). To address this, the external information was divided into more information sources, each of them promoting one of the possible choices. Here we analysed the situation where each choice is equally promoted, and allowed for modulation of each of the individual information sources. Results showed that multiple information sources are crucial for obtaining stable clusters that are not extreme, unlike the situation with one information source. Additionally, full agreement of the population to one information source was observed for cases where information was not extremely mild (hence the consensus does not imply indecision). Another important observation is the fact that, although all choices are equally promoted, differentiation can be seen in the population, i.e. each individual prefers one choice over the other. This was not the case for one information source, since an extremely mild information caused a generalised state of indecision in the population. These new features of the model, introduced by the existence of multiple external information sources, are important aspects for modelling many real situations. 

The model presented here can be further extended to use more information sources than number of opinions, to include variation in the individual interacting preferences and to account for a diluted interaction graph. Furthermore, application of the model to real data is planned, including environmental awareness with sensor data information and political opinions under the influence of newspapers.
 
\section*{Acknowledgments}
This research has been supported by the EveryAware project
funded by the Future and Emerging Technologies program (IST-FET) of
the European Commission under the EU RD contract IST-265432 and the
EuroUnderstanding Collaborative Research Projects DRUST funded by the
European Science Foundation.


\bibliographystyle{ws-acs}
\bibliography{refs}

\end{document}